\documentclass{ws-procs9x6}

\begin{document}

\title{Beyond the SM; D-brane model building from intersecting brane worlds}

\author{C.~Kokorelis}

\address{Institute of Nuclear Physics, N.C.S.R. Demokritos, GR-15310, Athens, Greece\\ 
}


\maketitle

\abstracts{
We discuss the appearance of non-supersymmetric D6-brane 
GUT model constructions. We focus 
on the construction of the first examples of flipped 
SU(5) and SU(5) GUTS which have only the SM at low energy. These 
constructions are based on
4D compactifications in $Z_3$ toroidal orientifolds of type IIA with 
D6-branes intersecting at angles. 
}


Intersecting brane worlds (IBW's) \cite{lu,iba,kokos1,ang,cve1,kokoiba,kokos5,kokos6,kokos7,igorwi,cve3,abel,loui,lust3,nano,cve4,kokos4,kokosnew,rev} have received a 
lot of attention recently as  it became possible on them to build 
toroidal D-brane models \footnote{in the absence of RR and NSNS fluxes} that have 
only the SM at low 
energy \cite{iba,kokos1,kokoiba,kokos5,kokos6,kokos7,kokos4}. One of those
SM constuctions may be reviewed in this talk.

Lets us review the intersecting D6-branes constructions of 
the $Z_3$ orientifolds of \cite{lust3}. The D6-branes satisfy
the following RR tadpole conditions, namely 
\footnote{
The net number of bifundamental massless chiral fermions in the models
is defined as
\begin{equation}
({\bar N}_a, N_b)_L :\  I_{ab} = Z_a Y_b - Y_a Z_b; \ 
(N_a, N_b)_L : \   I_{ab^{\star}} = Z_a Y_b + Y_a Z_b
\label{spec1}
\end{equation}
}
\begin{equation}
\sum_a N_a Z_a = 2 \ .
\label{tad}
 \end{equation}
As it was noticed in \cite{lust3} the simplest realization of an SU(5) GUT
involves two stacks of D6-branes at the string scale $M_s$, the
first one corresponding to a $U(5)$ gauge group while the second one to a
U(1) gauge group. Its effective wrapping numbers are given by 
$(Y_a,
Z_a) = (3, \frac{1}{2}), \  (Y_b, Z_b) = (3, -\frac{1}{2})
$. Under the decomposition $ U(5) \subset SU(5)
\times U(1)_a$, the models become effectively an $SU(5) \times
U(1)_a \otimes U(1)_b$ GUT. One combination of U(1)'s
become massive due to its coupling to a RR field, 
another one remains massless to low energies. The spectrum of this
SU(5) GUT may be seen in the first seven columns (reading from the left) 
of the following table. 

\begin{table}[htb]
{\footnotesize
\begin{tabular}{@{}cccccccc@{}}
\hline \hline
\multicolumn{8}{c}{}\\[-2ex]
Field & Sector & $\#$ & SU(5)    &  $U(1)_a$ & $U(1)_b$
 & $U(1)^{mas}$ & $U(1)^{fl} = \frac{5}{2} U(1)^{mas}$ \\
\multicolumn{8}{c}{}\\[-2ex]
f & $\{ 51 \}$ & 3  & ${\bf {\bar 5}}$ & $-1$ &  1   & $-\frac{6}{5}$ & $-3$    \\
F & $A_{a}$ & 3          & ${\bf 10}$   &  2         &   0 & $\frac{2}{5}$ &   1          \\
$l^c$ & $S_{b}$ & 3          &  ${\bf 1}$          &   0        &   -2   & 2 & 5    \\
  \hline
\end{tabular}
\label{table2}}
\end{table}
At this stage the SU(5) models - have the correct chiral 
fermion content of an SU(5) GUT - and the extra U(1) surviving 
the presence of the Green-Schwarz mechanism, 
breaks by the use of a singlet field present. However, 
the electroweak ${\bf }5$-plets needed for electroweak
 symmetry breaking of the models are absent. [Later on, attempts to 
construct a
fully N=1 supersymmetric SU(5) models at $M_s$ in \cite{cve4}, produced 
3G models
that were not free of remaining massless exotic 15-plets.] 
Also, later on 
in \cite{nano} it was noticed that if one leaves unbroken, 
and rescales, the U(1) 
surviving massless the Green-Schwarz mechanism of the SU(5) GUT of 
\cite{lust3}, the rescaled U(1) becomes
the flipped U(1) generator. However, the proposed 3G models lacked the presence
of GUT Higgses or electroweak pentaplets and were accompanied by extra
exotic massless matter to low energies. We note that the
 charges under the $U(1)^{fl}$ gauge symmetry,
 when rescaled
appropriately (and $U(1)^{fl}$ gets broken) `converts' the flipped SU(5) model
to a three
generation (3G) SU(5).

In \cite{kokos4} we have shown that it is possible to construct the first 
examples of string 
SU(5) and
 flipped
SU(5) GUTS - where we identified the appropriate GUT and electroweak 
Higgses - which break to the SM at low energy.
In the flipped SU(5) GUT, the fifteen fermions of the SM plus the right
handed neutrino $\nu^c$ belong to the 
\begin{equation}
F = {\bf 10_1} = (u, d, d^c, \nu^c), \ \
f = {\bf {\bar 5}_{-3}} = (u^c, \nu, e), \  \ l^c =  {\bf 1_{5}} = e^c
\label{def1}
\end{equation}
chiral multiplets.
The GUT breaking Higgses may come from the `massive' spectrum
of the sector localizing
the ${\bf 10}$-plet (${\bf 10_1^B} = (u_H, d_H, d^c_H, \nu^c_H )$ 
fermions seen in the table below. The lowest order
Higgs in this sector, let us call them $H_1$, $H_2$, have quantum numbers as
those given in the table below.
\begin{table} [htb] \footnotesize
\renewcommand{\arraystretch}{1}
\begin{center}
\begin{tabular}{|c|c|c|c|c|c|}
\hline
Intersection & GUT Higgses & repr. & $Q_a$ & $Q_b$ & $Q^{fl}$\\
\hline
$\{ a,{\tilde O6} \}$  & $H_1$  &  {\bf 10}   & $2$   & $0$ & $1$ \\
\hline
$ \{ a,{\tilde O6} \}$  & $H_2$  &  ${\bf {\bar 10}}$   & $-2$ & $0$  & $-1$ \\
\hline
\end{tabular}
\end{center}
\caption{\small 
Flipped $SU(5) \otimes U^{fl}$ GUT symmetry breaking
scalars. 
\label{Higgsfli}}
\end{table}
By looking at the last column of the table, we realize that the
Higgs $H_1$, $H_2$ are the GUT symmetry breaking Higgses of a standard
flipped SU(5) GUT.
By dublicating the analysis,
one may conclude that what it appears in the effective theory as GUT 
breaking Higgs scalars, is the combination 
$
H^G = H_1 + H_2^{\star}$.
In a similar way the correct identification
of the electroweak content \cite{kokos4} of the flipped SU(5) ${\bf 5_{-2}^B} = (D, h^{-},  h^{0} )$-plet
(and 
SU(5))GUTS made possible 
the existence of the see-saw mechanism which is generated by the 
interaction 
\begin{eqnarray}
L = {\tilde Y}^{\nu_L \nu_R} \cdot {\bf 10} \cdot {\bar {\bf
5}} \cdot {\bar h}_4 \
      + \  {\tilde Y}^{\nu_R} \cdot \frac{1}{M_s} \cdot ({\bf 10} \cdot
{\bf \overline{10}}^B)
      ({\bf 10} \cdot {\bf \overline{10}}^B)\ . 
\label{seesaw1}
\end{eqnarray}
Its standard version can be generated by choosing
\begin{equation}
\langle h_4 \rangle = \upsilon, \  \langle {\bf 10}_i^B \rangle = M_s
\label{masse1}
\end{equation}
and generates small neutrino masses.
In these constructions the baryon number is not a gauged symmetry, thus 
a high GUT scale of the order of the $10^{16}$ GeV helps the theory to
avoid gauge mediated proton decay modes like the \cite{kokos4}
\begin{equation}
  \sim \frac{1}{M_s^2}\ ({\bar u}^c_L \ u_L) \ ({\bar e}_{R}^{+}) 
(d_{R}),\ \
\sim \frac{1}{M_s^2}\ ({\bar d}^c_R \ u_R) ({\bar d}^c_L \ \nu_L) \ .
\end{equation}
[In IBW's proton decay by direct calculation of string amplitudes
for SUSY SU(5) D-brane models was examined in \cite{igorwi}.]
Also scalar mediated proton decay modes get suppressed by the existence of 
a new solution to the doublet-triplet splitting problem   
 \begin{equation}
\frac{r}{M_s^3}(HHh)( {\bar F}  {\bar F} {\bar h}) + m ({\bar h}h) (
{\bar H} H) + \kappa ({\bar H}H)( 
{\bar H} H),\end{equation}
that stabilizes the vev's of the triplet scalars $d_c^H$, $D$ \cite{kokos4}.
This is the first example of a doublet-triplet splitting realization
in IBW's. 
The full solution of the gauge hierarchy problem,
that is avoiding the existence of quadratic corrections to the electroweak 
Higgses remains an open issue in the present GUTS.  


%
%
%
%

\end{document}